\documentstyle[11pt]{article}
\input epsf
\def\sdtimes{\mathbin{\hbox{\hskip2pt\vrule
height 4.1pt depth -.3pt width .25pt\hskip-2pt$\times$}}}

\begin{document}
\thispagestyle{empty}
\baselineskip 20pt
\rightline{KIAS-P00020}
\rightline{SNUTP00-011}
\rightline{{\tt hep-th}/0005056}
\

\def\sdtimes{\mathbin{\hbox{\hskip2pt\vrule
height 4.1pt depth -.3pt width .25pt\hskip-2pt$\times$}}}

\def\tr{{\rm tr}\,}
\newcommand{\beq}{\begin{equation}}
\newcommand{\eeq}{\end{equation}}
\newcommand{\beqn}{\begin{eqnarray}}
\newcommand{\eeqn}{\end{eqnarray}}
\newcommand{\bde}{{\bf e}}
\newcommand{\balpha}{{\mbox{\boldmath $\alpha$}}}
\newcommand{\bsalpha}{{\mbox{\small\boldmath$alpha$}}}
\newcommand{\bbeta}{{\mbox{\boldmath $\beta$}}}
\newcommand{\btau}{{\mbox{\boldmath $\tau$}}}
\newcommand{\blambda}{{\mbox{\boldmath $\lambda$}}}
\newcommand{\bepsilon}{{\mbox{\boldmath $\epsilon$}}}
\newcommand{\bphi}{{\mbox{\boldmath $\phi$}}}
\newcommand{\bpi}{{\mbox{\boldmath $\pi$}}}
\newcommand{\bX}{{\mbox{\boldmath $X$}}}
\newcommand{\ggg}{{\boldmath \gamma}}
\newcommand{\ddd}{{\boldmath \delta}}
\newcommand{\mmm}{{\boldmath \mu}}
\newcommand{\nnn}{{\boldmath \nu}}

\newcommand{\bra}[1]{\langle {#1}|}
\newcommand{\ket}[1]{|{#1}\rangle}
\newcommand{\sn}{{\rm sn}}
\newcommand{\cn}{{\rm cn}}
\newcommand{\dn}{{\rm dn}}
\newcommand{\diag}{{\rm diag}}

\

\vskip 1cm

\centerline{\Large\bf General Dynamics of Two Distinct BPS Monopoles}

\vskip 1.5cm
\centerline{\large\it Choonkyu Lee$^{a}$\footnote{cklee@phya.snu.ac.kr}
and Kimyeong Lee$^{b}$\footnote{klee@kias.re.kr}}
\vskip 3mm
\centerline{$^a$Department of Physics and Center for Theoretical Physics}
\centerline{Seoul National University,  Seoul 151-742, Korea}
\vskip 3mm
\centerline{$^b$School of Physics, Korea Institute for Advanced Study}
\centerline{207-43, Cheongryangri-Dong, Dongdaemun-Gu}
\centerline{ Seoul 130-012, Korea}
\vskip 1.5cm
\vskip 5mm
\begin{quote}
{\baselineskip 16pt Classical and quantum dynamics of two distinct BPS
monopoles in the case of non-aligned Higgs fields are studied on the
basis of the recently determined low energy effective theory. Despite
the presence of a specific potential together with a kinetic term
provided by the metric of a Taub-NUT manifold, an O(4) or  O(3,1)
symmetry of the system allows for a group theoretical derivation of
the bound-state spectrum and the scattering cross section.}

\end{quote}

\newpage
\setcounter{footnote}{0}

Recently, much attention has been given to the low energy interaction
of BPS monopoles in the case that a semi-simple gauge group $G$ of
rank $k$ breaks to its maximal torus $U(1)^k$ by one or more Higgs
fields in the adjoint representation. In the simplest non-trivial case
of the rank two and when the Higgs expectation values have only one
independent component, the relative dynamics of two distinct
fundamental BPS monopoles can be described by the kinetic Lagrangian
provided by the metric of a Taub-NUT
manifold~\cite{rank2,metric}. Here, relevant variables consist of the
relative position ${\bf r}$ and the relative phase $\psi \in
[0,4\pi]$, and the Lagrangian takes the form
\beq
L_{\rm kin} = \frac{ \mu}{2}\left\{ \left( 1 + \frac{r_0}{r}\right)
\dot{\bf r}^2 + r_0^2 \left(1+ \frac{r_0}{r}\right)^{-1} (\dot{\psi} +
{\bf w}({\bf r})\cdot \dot{\bf r} )^2 \right\} ,
\eeq
where $\mu$ is the reduced mass, and ${\bf w}({\bf r}) $ is a Dirac
monopole potential satisfying $\nabla \times {\bf w} = - {\bf r}/r^3$.
The positive parameter $r_0$ of length dimension provides the length scale of
the problem.  The metric of the Taub-NUT space is chosen so that
$L_{\rm kin} = \mu g_{\mu\nu}\dot{x}^\mu \dot{x}^\nu /2$ with the
coordinates $x^\mu = ({\bf r}, r_0 \psi)$. On the other hand, if the
Higgs expectation values have more independent components, it has been
shown recently~\cite{klee} that the low energy Lagrangian acquires an
additional attractive potential term.  (For more than two distinct
monopoles, there can be more than one term.)  That is, the Lagrangian
is of the form
\beq
L = L_{\rm kin} - U(r)
\label{lag}
\eeq
with 
\beq
U(r) = \frac{1}{2\mu r_0^2} \frac{a^2}{1+ \frac{r_0}{r}},
\label{pot}
\eeq
where $a>0$ is a constant related to the degree of the vacuum
misalignment.

In this paper we will present the analysis on classical and quantum
dynamics of the system defined by the Lagrangian (\ref{lag}) and exhibit
physical consequences due to the potential (\ref{pot}). 

With $U(r)=0$ and the parameter $r_0$ taken to be negative, our system
reduces to the previously considered case in
Ref.~\cite{gibbons,ruback,feher} for an approximate description of the
relative dynamics concerning two identical BPS monopoles. In this case
the system was found to possess the O(4)/O(3,1) dynamical symmetry
analogous to the one in the well-known Coulomb problem, and this in
turn allowed an algebraic derivation of the dyonic bound state spectrum
and the two monopole scattering cross section as shown
in~Ref.~\cite{gibbons,ruback,feher}.  This analysis requires no
essential modification for our case, except that no dyonic bound
states become available when the potential $U(r)$ is absent.  In our
case the relative charge leads to a repulsion instead of an attraction
between two BPS monopoles.

What changes would the presence of the potential (\ref{pot}) in the
case of non-aligned Higgs fields bring?  What is  remarkable about
our system is that the O(4)/O(3,1) dynamical symmetry operates even in
the presence of this potential and so we can determine the effect due
to the potential exactly. Moreover, we will see that at the level of a
time-independent Schr\"odinger equation, our system shares the same form
as the Zwanziger system~\cite{zwan}-a model involving a suitably combined
monopole+ Coulomb +$ 1/r^2$ potential with  O(4)/O(3,1)
symmetry. Based on this observation, we can deduce the exact spectrum
of dyonic bound states (which exist only in the non-aligned Higgs
vacuum), and also the scattering cross section for two distinct BPS
monopoles in the general case.

We will study the classical dynamics first. Suppose that we have not
yet posited the precise form of a radial potential $U(r) $ in
Eq.~(\ref{lag}). The Lagrangian has two obvious conserved quantities
associated with the cyclic variable $\psi$ and the time $t$, viz.,
\beqn
&& q= \mu r_0^2 (1+ \frac{r_0}{r})^{-1} (\dot{\psi}+{\bf w}({\bf
r})\cdot \dot{\bf r}), \\
&& E= \frac{\mu}{2}(1+ \frac{r_0}{r}) \left[ \dot{\bf r}^2 +
\frac{q^2}{\mu^2 r_0^2} \right] + U(r) ,
\label{energy}
\eeqn
corresponding to the relative electric charge and energy,
respectively. The mechanical three momentum 
\beq
\bpi = \mu (1+ \frac{r_0}{ r}) \dot{\bf r}
\eeq
satisfies the equation of motion
\beq \dot{\bpi}= - \frac{\mu r_0}{2} \frac{{\bf r}}{r^3} \left(\dot{\bf
r}^2 - \frac{q^2 }{\mu^2 r_0^2} \right) -q\frac{\dot{\bf r}\times {\bf
r}}{r^3} - \hat{{\bf r}}\; U'(r).
\label{pieq}
\eeq
There is also the conserved angular momentum,
\beq
{\bf J} = {\bf r} \times \bpi + q \hat{\bf r},
\eeq
and so the position vector ${\bf r}$ moves on a cone defined by 
\beq
{\bf J}\cdot \hat{\bf r} = q.
\eeq
At this point, let us further demand, as a restriction on the
potential $U(r)$, that all bound trajectories in the system be closed
orbits. Such is the case if there is an additional conserved quantity
of the Runge-Lenz type, ${\bf K}=\bpi \times {\bf J} - \beta \hat{\bf
r}$ with a constant $\beta$. A simple calculation using
Eq.~(\ref{pieq}) determines the potential to be of the form
(\ref{pot}) up to a possible constant shift.  The precise form of the
conserved Runge-Lenz vector is also fixed as
\beq
{\bf K} = \bpi \times {\bf J} -  \mu r_0 \left( E- \frac{q^2}{\mu
r_0^2} \right) \hat{\bf r}.
\eeq
(Essentially, this corresponds to a generalization of the Bertrand
theorem~\cite{bert} to the case with the Taub-NUT kinetic term.) Our
potential thus assumes a very special status-it corresponds to the
general attractive potential which admits both the angular momentum
and the Runge-Lenz vector as conserved quantities.

One may wonder whether there exists any additional modification of the
Taub-NUT dynamics without jeopardizing the existence of a conserved
Runge-Lenz type vector. One such modification~\cite{iwai} is to change
the Taub-NUT metric to
\beq
ds^2 = \frac{a+br}{r} d{\bf r}^2 + \frac{ ar+ br^2}{1 + cr + dr^2}
(d\psi + {\bf w}({\bf r})\cdot d{\bf r})^2.
\eeq
For a particle moving in the space with this metric, there still
exists  a conserved Runge-Lenz vector~\cite{iwai}. However, the metric
is always singular somewhere near the origin. This modification is
rather ad hoc and does not have any good physics motivation. This
metric also has nonzero Riemann tensor $R_{\mu\nu}$ and so is not
hyperk\"ahler.  Further exploration of this modification was studied
in Ref.~\cite{iwai,visi}. Another possibility, which is somewhat
novel, is to add a term linear in velocity,
\beq
L_{\rm linear} = \frac{ c}{1+ \frac{r_0}{r} } (\dot{\psi} + {\bf w}({\bf
r})\cdot \dot{\bf r}) .
\eeq
This term can be absorbed by shifting $\psi$ to $\psi + c t/(\mu
r_0^2)$ and $a^2 \rightarrow a^2+ c^2$ in the Lagrangian
(\ref{lag}). Thus the physics does not change much due to this
modification. Thus, we will focus on our Lagrangian (\ref{lag}).

Even in the presence of the potential (\ref{pot}), a complete
description of the motion can now be given on the basis of the
conservation laws.  {}From two conserved quantities ${\bf J}$ and ${\bf
K}$, we get the conserved quantities
\beqn 
&& {\bf J}\cdot {\bf K} = - \mu r_0 q (E-\frac{q^2}{\mu r_0^2}),
\label{cla1}
\\ 
&& {\bf K}^2 = 2\mu ({\bf J}^2-q^2)(E-\frac{q^2+a^2}{2\mu r_0^2}) +\mu^2
r_0^2 (E-\frac{q^2}{\mu r_0^2})^2 .
\label{cla2}
 \eeqn
Defining a conserved vector
\beq
{\bf N} = q{\bf K} + \mu r_0 \left( E-\frac{q^2}{\mu r_0^2} \right) {\bf J}, 
\eeq
we also find that 
\beq
{\bf N}\cdot{\bf r} = q( {\bf J}^2-q^2).
\eeq
{}From this equation, it follows that the motion is confined on a plane
perpendicular to the constant vector ${\bf N}$. Thus the trajectories
belong to a plane intersecting the cone, i.e., correspond to conic
sections. The detailed orbit characteristics can be inferred from the
energy equation (\ref{energy}). {}From Eq.~(\ref{energy}), the effective
potential for a given $q$ is
\beq
U_{\rm eff}(r) = \frac{q^2}{2\mu r_0^2}(1+ \frac{r_0}{r}) +
\frac{1}{2\mu r_0^2}
\frac{a^2}{1+\frac{r_0}{r}}.
\eeq
For $|q|\ge a$, $U_{\rm eff}(r)$ is a monotonically decreasing function
of $r$ and so there is no bound state. In this case, the energy is
bounded by the minimum of the effective potential $U(\infty)$, i.e.,
$E\ge (q^2+a^2)/(2\mu r_0^2 )$. When the equality holds for this bound, the
trajectories are parabolic orbits, and otherwise we find the
hyperbolic orbits.  On the other hand, if the relative electric charge
satisfies $|q|<a$, the potential $U_{\rm eff}(r)$ assumes the minimum
value $E_{\rm bps}= a|q|/(\mu r_0^2)$ at finite radius $r_{\rm bps}=
r_0/(a/|q|-1)$.  Thus, with $|q|<a$, we have the following
possibilities,
\begin{description}
\item  {\hskip10mm} $E> \frac{1}{2\mu r_0^2}(q^2+a^2)$: (unbounded)
hyperbolic orbits,  
\item {\hskip10mm} $E=\frac{1}{2\mu r_0^2}(q^2+a^2)$: parabolic orbit,
\item {\hskip10mm} $\frac{1}{2\mu r_0^2}(q^2+a^2)>E>E_{\rm bps}$:
(bounded) elliptic orbits,
\item {\hskip10mm} $E=E_{\rm bps}$: a static configuration with
$\dot{\bf r}=0$,  $r=r_{\rm bps}$ and ${\bf J} = q \hat{\bf r}$. 
\end{description}
The last case where $E=E_{\rm bps}$ can be regarded as the classical
1/4 BPS configuration~\cite{yi,klee}.

For the scattering orbits, the two conserved quantities ${\bf J}$ and
${\bf K}$ can also be utilized in determining how the scattering angle
$\Theta$ depends on the impact parameter $b$, the initial relative
speed $v$ (or the initial energy $E$), and the relative charge
$q$. Since the detailed arguments on this can be found from
Ref.~\cite{gibbons}, we shall here give only the formula pertaining to
our case:
\beqn
\tan \frac{ {\Theta}}{2} &=& \frac{1}{b}\sqrt{ \frac{q^2}{\mu^2
v^2} + \frac{r_0^2}{\mu^2 v^4} \left(E - \frac{q^2}{\mu r_0^2}\right)^2} \\
&=&  \frac{r_0}{2b}\sqrt{
\left(1 + \frac{a^2+q^2}{\mu^2 r_0^2 v^2}\right)^2 - \frac{4a^2
q^2}{\mu^4
r_0^4 v^4} }\; ,
\eeqn
where we have used the relation 
\beq
E= \frac{1}{2}\mu v^2 + \frac{1}{2\mu r_0^2}(q^2+a^2)
\eeq
for the initial variables.  This leads to the Rutherford-type
differential cross section
\beqn \frac{d \sigma}{d \Omega} &=& \frac{b}{\sin{ \Theta}} \left|
\frac{ db}{d{\Theta}}\right| \\ &=& \frac{r_0^2 }{16}\left\{ \left(1 +
\frac{a^2+q^2}{\mu^2 r_0^2 v^2}\right)^2 - \frac{4a^2 q^2}{\mu^4 r_0^4
v^4} \right\} \csc^4 \frac{{ \Theta}}{2}\;.
\label{cross1}
\eeqn
Observe that when $a$ vanishes, this  reduces to the old
result~\cite{gibbons,ruback,feher}.

Let us now turn to quantum  dynamics. For the classical
Lagrangian (2), the canonical momentum conjugate to ${\bf r}$ is given
by 
\beq 
{\bf p }= \frac{ \partial L}{\partial \dot{\bf r}}= \bpi + q{\bf
w}({\bf r}),
\eeq
and the Hamiltonian by
\beq
H = \frac{1}{2\mu(1+ r_0/r)} ( \bpi^2 + \frac{a^2}{r^2_0}) +
\frac{1}{2\mu} (1+ \frac{r_0}{r})\frac{ q^2}{r_0^2}.
\label{hamil}
\eeq
In the quantum theory, dynamical variables become operators and so,
for an expression involving operators like our Hamiltonian, an
appropriate ordering must be specified. We will here follow the
standard procedure developed for a general nonlinear
system~\cite{dewitt}. The variables $x^\mu = ({\bf r}, r_0 \psi)$ and their
conjugates $p_\mu= ({\bf p},q/r_0)$ are taken to be hermitian operators
with respect to the Taub-NUT volume measure $\sqrt{g}d^4 x = (1+
r_0/r)d^4 x $ with $g={\rm det}(g_{\mu\nu})$, and satisfy the basic
commutation relations $[x^\mu,x^\nu] = 0, [p_\mu,p_\nu]=0$ and
$[x^\mu, p_\nu] = i\hbar \delta^\mu_\nu$. ({}From now on, we put
$\hbar=1$.) In this representation, the canonical momentum operator is
\beq
p_\mu = -i (\partial_\mu + \frac{1}{4}g^{-1}\partial_\mu g).
\eeq
The Hamiltonian operator is then identified with~\cite{dewitt}
\beqn
H &=& \frac{1}{2\mu} g^{-1/4}p_\mu \sqrt{g} g^{\mu\nu} p_\nu g^{-1/4} +
\frac{1}{2\mu r_0^2} \frac{a^2}{1+ \frac{r_0}{r}}\\
&=& -\frac{1}{2\mu}
\frac{1}{\sqrt{g}}\partial_\mu(\sqrt{g}g^{\mu\nu}\partial_\nu) + 
\frac{1}{2\mu r_0^2} \frac{a^2}{1+ \frac{r_0}{r}} . 
\label{hamil1}
\eeqn
The relative charge operator $q=r_0 p_4 = -i \partial/\partial \psi$
commutes with the Hamiltonian and has eigenvalues $ s$ with $s=0,
\pm 1/2, \pm 1,...$. If one also introduces a non-hermitian operator
\beq
\bpi = -i \nabla - q {\bf w}({\bf r}),
\label{bpi}
\eeq
which satisfies the commutation relations
\beq
[\pi_i, x^j] = -i  \delta_i^j,\;\; [\bpi, \psi] = i {\bf w},\;\;
[\pi_i, \pi_j] = -i q\epsilon_{ijk} \frac{x^k}{r^3},
\eeq
it is a matter of lengthy computations~\cite{feher} to show that the
Hamiltonian (\ref{hamil1}) can be recast precisely into the form
(\ref{hamil}), now with the operator ordering taken into
consideration. The Hamiltonian operator in Eq.~(\ref{hamil}) is
hermitian with respect to the Taub-NUT volume measure. 

We are interested in the solution of the time-independent
Schr\"odinger equation, $H\Psi = E\Psi$. For the eigenstates with 
$q= s$, we use the Hamiltonian (\ref{hamil}) and factor out the
trivial phase $e^{iq\psi}$ from the wave function $\Psi$ to 
obtain
\beq
\left\{ \frac{1}{2\mu}  \bpi^2 - \frac{\alpha}{ r} + \frac{q^2}{2\mu
r^2} - {\cal E} \right\} \Psi = 0 ,
\label{dan}
\eeq
where two new constants $\alpha$ and ${\cal E}$ are given by
\beq
\alpha = r_0 (E- \frac{q^2}{\mu r_0^2}), \;\;\;\; {\cal E} = E -
\frac{1}{2\mu r_0^2 }  (a^2+q^2) .
\label{rela}
\eeq
Disregarding the above relations,  one may look upon Eq.~(\ref{dan}) as
the Schr\"odinger equation of a flat-space system, ${\cal H}\Psi = {\cal
E}\Psi$, with the Hamiltonian,
\beq {\cal H} = \frac{1}{2\mu} \bpi^2 - \frac{\alpha}{ r} +
\frac{q^2}{2\mu r^2} .
\label{dan1}
\eeq
This Hamiltonian is hermitian with respect to the usual Euclidean
three space volume measure and contains, aside from the magnetic
monopole and Coulomb-type potentials, a $1/r^2$ repulsive potential of
a definite strength. The very system was first considered by
Zwanziger~\cite{zwan}; Eq.~(\ref{dan1}) describes the relative
dynamics in his exactly soluble nonrelativistic model of two
dyons. It enjoys the $O(4)/O(3,1)$ symmetry. Our system also enjoys
the $O(4)/O(3,1)$ symmetry, which was noticed earlier without the
potential~(\ref{pot})~\cite{ruback,feher}.  In our case the symmetry
generators are quantum operators corresponding to the angular momentum
and the Runge-Lenz vector:
\beqn
&& {\bf J} = {\bf r}\times  \bpi + q \hat{\bf r} , \\
&& {\bf K} = \frac{1}{2} (\bpi \times {\bf J} - {\bf J} \times \bpi) 
- \mu r_0 \hat{\bf r} (H- \frac{q^2}{\mu r_0^2 })  ,
\eeqn
where $H$ is our full Hamiltonian operator (\ref{hamil}). (Note that
the order of various operators is important now.) Then by a bit of
lengthy algebra, one can verify that they satisfy the following
relations:
\beqn
&& [{\bf J}, H] = [{\bf K}, H] = 0 ,\\
&& [J_i, J_j]= i\epsilon_{ijk} J_k ,\\
&& [J_i, K_j]= i\epsilon_{ijk}K_k , \\
&& [K_i, K_j] = -i\epsilon_{ijk} J_k (2\mu H - \frac{q^2+a^2}{r_0^2}) .
\eeqn
They also satisfy 
\beqn
&& {\bf K}\cdot {\bf J}={\bf J}\cdot {\bf K}= -\mu r_0 q (H-
\frac{q^2}{\mu r_0^2}) , 
\label{cla3} \\
&& {\bf K}^2 - \mu^2 r_0^2 (H-\frac{q^2}{\mu r_0^2})^2 = 2\mu ({\bf J}^2
- q^2 + \hbar^2)( H- \frac{q^2+a^2}{2\mu r_0^2}) .
\label{cla4}
\eeqn
With the explicit $\hbar$ correction in the quantum expression term,
one can compare these equations with their classical
counterparts~(\ref{cla1}) and (\ref{cla2}).  For the Zwanziger system,
the corresponding $O(4)/O(3,1)$ generators are made of ${\bf J}$ and
\beq
\tilde{\bf K} = \frac{1}{2} (\bpi \times {\bf J} - {\bf J} \times \bpi) 
- \mu \alpha \hat{\bf r}  ,
\eeq
which is obtained from ${\bf K}$ by replacing the operator
$r_0(H-q^2/\mu r_0^2)$ 
with the parameter $\alpha$ of Eq.~(\ref{rela}). They commute with
${\cal H}$ and satisfy the similar algebra with $[\tilde{K}_i,
\tilde{K}_j] = -i\epsilon_{ijk} J_k (2\mu {\cal H})$, and so on. The
analogue of Eqs.~(\ref{cla3}) and (\ref{cla4}) for these operators are
\beqn
&& \tilde{\bf K}\cdot {\bf J} = {\bf J}\cdot \tilde{\bf K}= -\mu \alpha q 
\label{cla5} \\
&& \tilde{\bf K}^2 -\mu^2\alpha^2 = 2\mu ({\bf J}^2 - q^2 + \hbar^2)
{\cal H}
\label{cla6}
\eeqn

As in the Coulomb system~\cite{pauli}, the bound state spectrum and
the scattering cross section for the Zwanziger system can be found
exactly with the help of an appropriate group representation
theory. Then, they may be recast  using the connection~(\ref{rela}), to
obtain the corresponding results for  our original system with the
Hamiltonian (\ref{hamil}). Of course, it is possible to use the group
representation theory directly with our system also. But the
procedure just mentioned is simpler. 

For $\alpha>0$, the Zwanziger system allows bound states with negative
energy ${\cal E}<0$. In this case, the operator ${\bf M} = \tilde{\bf
K}/\sqrt{2\mu | {\cal E}|} $ and ${\bf J}$ satisfy an $O(4)$
algebra. Introduce 
${\bf A} = ({\bf J} + {\bf M})/2$ and ${\bf B} = ({\bf J} -
{\bf M})/2$ as the generators for two $O(3)'s$ of $O(4) =
O(3)\times O(3)$. Then an energy eigenstate $\Psi $ may be taken as
$|a, a_3, b, b_3>$ with  integers or half-integers for $a,a_3$ and
$b,b_3$, satisfying
\beqn
&& {\bf A}^2 \Psi = a(a+1)\Psi, \;\; A_3 \Psi = a_3 \Psi, \\
&& {\bf B}^2 \Psi = b(b+1)\Psi, \;\; B_3 \Psi = b_3 \Psi .
\eeqn
Then, making a judicious use of the identities~(\ref{cla5}) and
(\ref{cla6}), it is not difficult~\cite{zwan} to derive the exact
bound state spectrum ${\cal E} = -\frac{ \mu \alpha^2}{2n^2}$, where
$n$ is the principal quantum number restricted to take values $n=
|s|+1,|s|+2,...$ if the eigenvalue $s$ of the  operator $q$ assumes
half-integers or integers.  This bound state energy levels are highly
degenerate, since, for  given $s$ and $n$, the quantum number $j$
associated with ${\bf J}^2 = ({\bf A}+ {\bf B})^2$ can run from $|s|$
to $n-1$. We can now translate these findings into those for our
original system. First of all, the conditions to allow bound states,
$\alpha>0$ and ${\cal E}<0$, translate into
\beq
\frac{s^2}{\mu r_0^2}< E < \frac{1}{2\mu r_0^2}(a^2+s^2),
\eeq
and so our system can have bound states only with $|s|< a$ as in the
classical case. On the other hand, inserting the formula ${\cal E} = 
-\frac{\mu \alpha^2}{2n^2}$ to Eq.~(\ref{rela}), we obtain a quadratic
equation for $E$,
\beq
-\frac{\mu r_0^2}{2n^2} (E-\frac{s^2}{\mu r_0^2})^2 = E-\frac{1}{2\mu
r_0^2 } (a^2+s^2)  .
\eeq
Solving this equation, we immediately obtain the desired bound state
spectrum for our system:
\beq
E = \frac{1}{\mu r_0^2}\left\{ -(n^2-s^2) + n \sqrt{n^2-s^2 +
a^2}  
\right\}
\label{energy1}
\eeq
with $|s|< a$ and $n\ge |s|+1$. The degeneracy of each energy
eigenstates is $n^2-s^2$ as $j$ runs from $|s|$ to $n-1$.

The case  with ${\cal E}>0$ corresponds to a scattering state. We
may put ${\cal E}= \mu v^2/2$ with the initial speed $v$. The
operators ${\bf M}$ and ${\bf J}$ now satisfy the O(3,1) algebra. As in
Ref.~\cite{zwan}, the group theory can be utilized to derive the exact
quantum mechanical scattering cross section formula for the Zwanziger
system, 
\beq
\frac{d \sigma}{d \Omega} =\frac{1}{4\mu^2v^4} ( s^2v^2+ \alpha^2)
\csc^4 \frac{\Theta}{2}.
\eeq
Again we translate this result into our case by using the relation
(\ref{rela}) and $q= s$ and see that this coincides with the classical
cross section formula~(\ref{cross1}), making the classical result to
be quantum mechanically exact.

For a given value of  $s$, the bound state energy spectrum in
Eq.~(\ref{energy1}) is an increasing function of $n$, satisfying the
inequality
\beq
E(n=|s|+1)\le E< E(n=\infty)=\frac{1}{2\mu r_0^2}(a^2+s^2).
\eeq
Especially, the ground state energy $E(n=|s|+1)$ with degeneracy
$2|s|+1$ is greater than the classical BPS energy $E_{bps}= a|s|/(\mu
r_0^2)$, which is identical to $E(n=|s|)$ if $n$ could take the value
$|s|$.  The difference between the ground state energy and the
classical minimum is due to the zero point energy of the bosonic
theory. It would be interesting to formulate the most general
supersymmetric Lagrangian ~\cite{klee,gaunt,dhoker} for our system and
find BPS and non-BPS bound states.  There would be no such energy
difference in supersymmetric vacua.

Finally, there exists an even larger conformal symmetry group
$O(4,2)/O(5,1)$ for the Taub-NUT system~\cite{ruback,feher}. This can
be easily generalized to our system, which we do not discuss here.
They will again play a role in the supersymmetric extension of our
work.

\centerline{\bf Acknowledgment}

We acknowledge useful discussions with D. Bak and P. Yi.  The work of
CL was supported in part by Korea Research Foundation
(1998-015-D00054) and the BK21 project of the Ministry of Education,
Korea and by the KOSEF Grant 97-07-02-02-01-3.  The work of KL was
supported in part by KOSEF 1998 Interdisciplinary Research Grant
98-07-02-01-5.

\appendix

\end{document}